\documentclass[12pt,titlepage]{article}
\usepackage{indentfirst}
\usepackage{psfrag}
\usepackage[dvips]{graphicx}
\usepackage{amsmath}
\usepackage{here}
\newcommand{\ptau}{\mbox{${\cal P}_{\tau}$}}

\newcommand{\ap}{\mbox{$A_{\cal P}$}}

\newcommand{\process}{\mbox{$e^+ e^- \rightarrow \tau^+ \tau^-$}}
\newcommand{\etal}{{\it et al.}}

\title{Measurement of $\tau$ polarization in $e^+ e^-$ annihilation at
$\sqrt{s}$=58 GeV}

\author{}
\date{}
\begin{document}


\small
\begin{flushright}
KEK Preprint 96-171\\
KOBE HEP 97-01\\
NGTHEP 97-01\\
OULNS 96-05\\
TMU HEP 97-20\\
\end{flushright}

\vspace{20mm}

\begin{center}
\LARGE{Measurement of $\tau$ polarization\\ 
in $e^+ e^-$ annihilation at $\sqrt{s}$=58 GeV}

\small 
\vspace{20mm}
H.~Hanai \etal   \\
VENUS Collaboration\end{center}

\vspace{40mm}

The polarization of $\tau$ leptons in the reaction $\process$
has been measured using a $e^+e^-$ collider, TRISTAN, at the center-of-mass
energy of 58 GeV.
From the kinematical distributions of daughter particles in
$\tau\rightarrow e\nu\bar{\nu}$, $\mu\nu\bar{\nu}$, $\rho\nu$ or $\pi(K)\nu$ decays,
the average polarization of $\tau^{-}$ and its forward-backward asymmetry
have been evaluated to be 0.012$\pm$0.058 and 0.029$\pm$0.057, respectively.

\begin{center}
\vspace{30mm}
(Submitted to Phys. Lett. B)
\end{center}
\newpage

\begin{center}VENUS Collaboration\end{center}


\noindent H.~Hanai$^{1,a}$, J.~Haba$^{2}$, 
K.~Abe$^3$,
K.~Amako$^2$, Y.~Arai$^2$, T.~Arima$^{4,b}$, Y.~Asano$^4$, M.~Chiba$^5$, Y.~Chiba$^6$,
M.~Daigo$^7$, M.~Fukawa$^{2,c}$, Y.~Fukushima$^{2}$, 
H.~Hamasaki$^{4}$, 
Y.~Hemmi$^{8}$, M.~Higuchi$^{9}$, T.~Hirose$^{5}$, Y.~Homma$^{10}$, N.~Hosoda$^{5,d}$, 
N.~Ishihara$^{2}$, Y.~Iwata$^{11}$, J.~Kanzaki$^{2}$, R.~Kikuchi$^{8}$,
T.~Kondo$^{2}$, T.~T.~Korhonen$^{2,12,e}$, H.~Kurashige$^{8}$, 
E.~K.~Matsuda$^{13}$, T.~Matsui$^{2}$, M.~Miura$^{4,f}$, K.~Miyake$^{8}$, S.~Mori$^{4}$,
Y.~Nagashima$^{1}$, Y.~Nakagawa$^{14,g}$, T.~Nakamura$^{15,h}$,
I.~Nakano$^{16,i}$, S.~Odaka$^2$, K.~Ogawa$^{2,h}$, T.~Ohama$^{2}$, T.~Ohsugi$^{11}$, H.~Ohyama$^{17}$,
K.~Okabe$^{13}$, A.~Okamoto$^{8}$, A.~Ono$^{18}$, J.~Pennanen$^{2,12}$,
H.~Sakamoto$^{8}$, M.~Sakuda$^{2}$, M.~Sato$^{9}$, 
N.~Sato$^{2}$, M.~Shioden$^{19}$, J.~Shirai$^2$, T.~Sumiyoshi$^{2}$, Y.~Takada$^{4}$,
-F.~Takasaki$^{2}$, M.~Takita$^{1}$, N.~Tamura$^{13,j}$, D.~Tatsumi$^{1}$,
K.~Tobimatsu$^{20}$, T.~Tsuboyama$^2$, S.~Uehara$^{2}$, Y.~Unno$^{2}$, T.~Watanabe$^{21}$, 
Y.~Watase$^{2}$, F.~Yabuki$^{5}$, Y.~Yamada$^{2}$, T.~Yamagata$^{14}$,
Y.~Yonezawa$^{22}$, H.~Yoshida$^{23}$ and K.~Yusa$^{4}$ 

\vspace{5mm}

\noindent$^1$Department of Physics, Osaka University, Toyonaka 560, Japan

\noindent$^2$KEK, National Laboratory for High Energy Physics, Tsukuba 305, Japan

\noindent$^3$Department of Physics, Tohoku University, Sendai 980, Japan

\noindent$^4$Institute of Applied Physics, University of Tsukuba, Tsukuba 305, Japan

\noindent$^5$Department of Physics, Tokyo Metropolitan University, Hachioji 192-03, Japan

\noindent$^6$Yasuda Women's Junior College, Hiroshima 731-01, Japan

\noindent$^7$Faculty of Economics, Toyama University,Toyama 930, Japan

\noindent$^8$Department of Physics, Kyoto University, Kyoto 606, Japan

\noindent$^9$Department of Applied Physics, Tohoku-Gakuin University, Tagajo 985, Japan

\noindent$^{10}$Faculty of Engineering, Kobe University , Kobe 657, Japan

\noindent$^{11}$Department of Physics, Hiroshima University, Higashi-Hiroshima 724, Japan

\noindent$^{12}$Research Institute for High Energy Physics, Helsinki University, 
SF-00170 Helsinki, Finland

\noindent$^{13}$Department of Physics, Okayama University, Okayama 700, Japan

\noindent$^{14}$International Christian University, Mitaka 181, Japan

\noindent$^{15}$Faculty of Engineering, Miyazaki University, Miyazaki 889-01, Japan

\noindent$^{16}$Institute of Physics, University of Tsukuba, Tsukuba 305, Japan

\noindent$^{17}$Hiroshima National College of Maritime Technology, Higashino 725-02, Japan

\noindent$^{18}$Faculty of Cross-Caltural Studies, Kobe University, Kobe 657, Japan

\noindent$^{19}$Ibaraki College of Technology, Katsuta 312, Japan

\noindent$^{20}$Center for Information Science, Kogakuin University, Tokyo 163-91, Japan

\noindent$^{21}$Department of Physics, Kogakuin University, Hachioji 192, Japan

\noindent$^{22}$Tsukuba College of Technology, Tsukuba 305, Japan

\noindent$^{23}$Naruto University of Education, Naruto 772, Japan

\vspace{5mm}

\noindent$^a$Present address: ASTEC Inc., BR Ichigaya, 6 Minami-cho,
Shinjuku-ku, Tokyo 162, Japan.

\noindent$^b$Present address: Faculty of Engineering, Kyushu University,
Fukuoka 812, Japan.

\noindent$^c$Present address: Naruto University of Education, Naruto 772,
Japan.

\noindent$^d$Present address: Japan Synchrotron Radiation Research Institute,
Kamigouri 678-12, Japan.

\noindent$^e$Present address: Accelerator Division, KEK, Tsukuba 305, Japan.

\noindent$^f$Present address: Institute for Cosmic Ray Research,
University of Tokyo, Tanashi 188, Japan.

\noindent$^g$Present address:Department of Mathematical Sciences, Ehime University, 
Matsuyama 790-77, Japan.

\noindent$^h$Deceased.

\noindent$^i$Present address: Department of Physics, Okayama University,
Okayama 700, Japan.

\noindent$^j$Present address: Department of Physics, Niigata University,
Niigata 950-21, Japan.

\newpage

\normalfont

\section{Introduction}

The polarization of
$\tau$ leptons in the reaction,
\begin{align}
	e^{+}e^{-} \rightarrow \tau^{+}\tau^{-}
\label{reaction}
\end{align}
\noindent provides us with information concerning the properties of the neutral
current of leptons \cite{jadach}. The measurement of it is, therefore,
important in every $Q^2$ range of the reaction.

The average polarization of $\tau$ leptons is defined as
\begin{align}
 \ptau = \frac{\sigma(h=+1)-\sigma(h=-1)}{\sigma(h=+1)+\sigma(h=-1)},
\end{align}

\noindent where $h(=\pm 1)$ is the helicity of $\tau^-$.
 Another relevant observable is the forward-backward asymmetry
of the polarization defined as
\begin{align}
 \ap = \frac{\sigma_F {\cal P}_F - \sigma_B {\cal P}_B}{\sigma_F + \sigma_B},
\label{ap}
\end{align}
\noindent where ${\cal P}_F$ and ${\cal P}_B$ are the average polarizations
measured in the forward and backward regions, respectively, relative to the incoming $e^{-}$ direction.

Although there are many precise measurements on  $\ptau$ and $\ap$ at
$Z^0$ pole \cite{LEP exp}, very few are available below
the pole \cite{below Z exp}.
Besides, the previous measurements below
the pole suffered from relatively large statistical errors due to the small
cross section of reaction (\ref{reaction}).
 
This letter
describes a measurement of $\ptau$ and $\ap$ at a center-of-mass energy ($\sqrt{s}$) of 58 GeV using the VENUS detector at TRISTAN.
The analysis is based on the data corresponding to an integrated luminosity of 271 pb$^{-1}$.

In the framework of the Standard Model \cite{WS}, the differential cross section of
the reaction~(\ref{reaction}) can be expressed for unpolarized electron and positron as
\begin{align}
 \frac{d\sigma}{d\Omega}(\cos\theta,h) & = 
 \frac{\alpha^2}{8s}\left[F_0(s)(1+\cos^2\theta)
 + 2F_1(s)\cos\theta
 - h\{F_2(s)(1+\cos^2\theta)
 + 2F_3(s)\cos\theta\}\right],
\label{cs}
\end{align}
where $\theta$ is the
scattering angle between $e^-$ and $\tau^-$.
The four form-factors are given as 
\begin{align}
 F_0(s) &= 1 + 2g_{V e}g_{V \tau}{\rm Re}\chi(s) +
  (g_{V e}^2 + g_{A e}^2)(g_{V \tau}^2 + g_{A \tau}^2)|\chi(s)|^2, \nonumber \\
 F_1(s) &= 2g_{A e} g_{A \tau}{\rm Re}\chi(s) +
  4g_{V e}g_{A e}g_{V \tau} g_{A \tau}|\chi(s)|^2, \nonumber \\
 F_2(s) &= 2g_{V e}g_{A \tau}{\rm Re}\chi(s) +
  2(g_{V e}^2 + g_{A e}^2)g_{V \tau}g_{A \tau}|\chi(s)|^2 \nonumber, \\
 F_3(s) &= 2g_{A e}g_{V \tau}{\rm Re}\chi(s) +
  2g_{V e}g_{A e}(g_{V \tau}^2 + g_{A \tau}^2)|\chi(s)|^2.
\label{F definition}
\end{align}

\noindent Here, $g_{V l}$ and $g_{A l}$ are the vector and axial-vector coupling constants, respectively, of
leptons $l$ to $Z^0$. The function $\chi(s)$ can be written using the mass of $Z^0$ ($M_Z$), its width ($\Gamma_Z$)  
and the weak mixing angle ($\theta_W$) as
\begin{align}
\chi(s)=\frac{1}{4\sin^2\theta_W\cos^2\theta_{W}}\times\frac{s}{s-M_Z^2+is\Gamma_Z/M_Z}.
\end{align}

\noindent Integrating Eq.~(\ref{cs}) over the full solid angle, the polarization and its forward-backward asymmetry
can be expressed as 
\begin{eqnarray}
{\cal P}_{\tau}&=&-\frac{F_2}{F_0}, \nonumber \\
A_{\cal P}&=&-\frac{3}{4}\frac{F_3}{F_0}.
\end{eqnarray}

Since the term including $|\chi(s)|$ are small at the TRISTAN energy, $\ptau$ is
sensitive to $g_{V e} g_{A \tau}$, and $\ap$ to $g_{A e} g_{V \tau}$.
On the other hand, they are sensitive to $A_{\tau}$ and $A_e$, respectively
at the $Z^0$ pole, where $A_l$ is defined as
${\cal A}_{l} \equiv 2g_{V l}g_{A l}/ (g_{V l}^2+g_{A l}^2)$.
In this sense, the present measurement is
complementary to those from the experiments at the pole.

\section{The VENUS Detector}
Since the VENUS detector is described in detail elsewhere \cite{VENUS-det}, we briefly
summarize those components relevant to the present measurement. The
central drift chamber (CDC) is the main component for charged particle tracks, located in a uniform magnetic field of
0.75 T parallel to the beam axis produced by a superconducting
solenoid. Charged particles with $|\cos\theta|\leq0.75$ ($\theta$ is
the angle with respect to the beam axis) are
detected in the CDC with a momentum resolution of
$\sigma_{p}/p\approx0.008p_t$(GeV/$c$), where $p_t$ is the transverse momentum
with respect to the beam axis. The angular resolution is
$8\sin^{2}\theta$ mrad and 1 mrad for 
polar and azimuthal angles, respectively. A large cylindrical
transition radiation detector (TRD) surrounds the CDC and covers an
angular region of $|\cos\theta|\leq0.68$. It provides an $e/\pi$ discrimination
capability as demonstrated in Fig.~1. 

Between the solenoid coil and the TRD, 96 time-of-flight (TOF)
counters are located at a radius
of 1.66 m. A time resolution of 200 ps is obtained for charged
particles within $|\cos\theta|\leq0.78$.
A cylindrical array of 5160 lead-glass blocks (LG) is
located outside the solenoid in order to measure the energy of
electrons and photons within $|\cos\theta|\leq0.80$. Its good energy
resolution (3.8\% for 30GeV electrons) makes it possible to identify electrons and $\pi^0$ clearly.
 Outside the
iron return yoke, eight layers of muon chambers (MU) are stacked, of
which six inner layers are aligned along the beam direction, while the
outermost two layers are stacked orthogonal to the inner ones.
 In the present analysis, the inner six layers are used for muon
identification in an angular range of $|\cos\theta|\leq0.5$.

\section{Selection of $\tau$ decays}
 The selection of the $\tau$ decay of a specific mode
proceeded in two steps. First, an preselection was made to reject non-$\tau$
events such as two-photon collision ($\gamma\gamma$ events) and multihadron
events. The $\tau$ pair events are characterized with such topology as
low-multiplicity, back-to-back narrow jets. The
criteria of the preselection are as follows:

\noindent 1) The number of CDC tracks should be between two and eight, where only those tracks with $p_t$ greater than 0.2 GeV/$c$ coming from the interaction region are counted inside $|\cos\theta|\leq0.50$.

\noindent 2) The total visible energy, the sum of the track momenta and the calorimeter energies must be greater than 0.2$\sqrt{s}$.

\noindent 3) The total energy deposited in the calorimeters must be
smaller than 0.8$\sqrt{s}$.

\noindent 4) The invariant mass calculated inside each of two hemispheres with respect to
the thrust axis must be smaller than 4.0 GeV$/c^2$.

\noindent 5) The acollinearity angle between the jet axes, defined by the vector sum of track momenta and energy flows in each
hemisphere must be smaller than 40$^{\circ}$.

\noindent The conditions were effective to reject  $\gamma\gamma$ events (2 and 5), Bhabha events (3) and multihadronic events (4).

In the second step, decay modes into a single-prong final state were
identified independently in each hemisphere. 
The decay modes considered here are
$\tau\rightarrow e\nu\bar{\nu}$, $\mu\nu\bar{\nu}$, $\pi(K)\nu$ and $\rho\nu$,
the sum of which amounts to 75 \% of all $\tau$ decays.

\begin{description}
\item [$\tau\rightarrow \mu\nu\bar{\nu}$ : ]
The muons were identified by using the MU.
The tracks which were associated with at least four corresponding hit layers of the MU
were identified as $\mu$. Those events in which both hemispheres were identified as 
$\mu\nu\bar{\nu}$ were discarded to reject $ee\rightarrow \mu\mu$ and $ee\mu\mu$ events.
\end{description}

\begin{description}
\item [$\tau\rightarrow\rho\nu$ : ]
For the non-$\mu$ tracks, the energy deposit in the TRD ($E_{\rm TRD}$) was required to be smaller than 10 keV to ensure the track to be $\pi^\pm$. Those tracks which were associated with one or two
neutral energy clusters in the LG inside a 30$^{\circ}$ cone were
taken as candidates of $\rho$.  Figure~2a) shows the invariant mass distribution of the selected two clusters. A peak corresponding to $\pi^0$ is clearly seen above small background. We selected those pairs whose invariant mass lies within 35
MeV/$c^2$ around the $\pi^0$ mass as candidates.
 Any single isolated cluster was taken as $\pi^0$ whose decay gammas
merged into one cluster. 
Finally, the invariant mass of the
$\pi^\pm$ and the $\pi^0$ candidates were calculated. Its distribution is shown in Fig.~2b). We selected those which lie within 230
MeV/$c^2$  around the $\rho$ mass as $\rho\nu$ decay candidates.

\end{description}

\begin{description}
\item [$\tau\rightarrow e\nu\bar{\nu}$ : ]
The electrons were identified by using the LG  and TRD information
 \cite{krueger}.  Among the tracks not selected in the preceding identifications, the candidate
tracks were selected by requiring 
$E/p$, the ratio of energy deposit in the LG to the track momentum, to be
greater than 0.8. $E_{\rm TRD}$ was also required to
exceed 10 keV to assure the passage of electron. Those events in which both hemispheres were
identified as a $e\nu\bar{\nu}$ decay were removed in order to reject Bhabha events and
$ee \rightarrow eeee$ events.
\end{description}

\begin{description}
\item [$\tau\rightarrow\pi(K)\nu$ : ]
Among those tracks which were associated with no substantial MU hits (one hit layer was allowed), $\pi^\pm$ tracks were 
selected by requiring $E/p < 0.8$ and $E_{\rm TRD}<$10 keV.
They were also required to have no neutral clusters with
energy greater than 200 MeV within 60$^{\circ}$ around them. Since $\pi/K$ separation was not possible in the present analysis, $\tau\rightarrow K\nu$ mode was also included in this category.
\end{description}
\noindent
Application of the TRD, which is one of the unique features of the VENUS
detector, assures reliable identification among $e\nu\bar{\nu}$, $\rho\nu\bar{\nu}$
and $\pi\nu$ decay modes.  

The number of the selected decays and the 
estimated background are summarized in Table 1.  
The efficiencies and contaminations for the above identification procedures
were evaluated as a function of the daughter particle momentum, using real data from other processes:  $ee\rightarrow ee$, $ee\gamma$ and $eeee$ as the electron
sample; $ee\rightarrow\mu\mu$, $\mu\mu\gamma$, $ee\mu\mu$ and cosmic rays
as the muon sample; $ee\rightarrow ee\pi\pi$, $\tau\rightarrow 3\pi\nu$
and $K_s\rightarrow 2\pi$ as the pion sample. It was found that there
were no apparent momentum dependence in the efficiencies nor contaminations.
They are also summarized in Table 1. For a confirmation, the branching ratios evaluated from our data are also listed together with the world average \cite{PDG}.
Those are in good agreements with one another.

\section{Evaluation of the polarization and its asymmetry}

In the case of the three body decays of $\tau$ leptons such as 
$\tau\rightarrow e\nu\bar{\nu}$ or $\mu\nu\bar{\nu}$, the energy spectrum of the 
charged leptons is expressed as \cite{tsai}
\begin{align}
 \frac{df}{dx} = a(x) + \ptau b(x),
 \label{dgdx3}
\end{align}
where $a(x)=(5-9x^2+4x^3)/3$, $b(x)=(1-9x^2+8x^3)/3$ and $x$ is the laboratory energy of the charged lepton scaled to the beam energy.
Figure~3 shows the observed $x$ distributions for a)$e\nu\bar{\nu}$ and b)$\mu\nu\bar{\nu}$ modes.

For the two body decays such as $\tau\rightarrow\pi\nu$ or
$\rho\nu$, the momentum spectrum of the decay particle is expressed in the same way as \cite{tsai}
\begin{align}
 \frac{df}{dx} = 1 + \alpha\ptau(2x-1)
 \label{dgdx2}
\end{align}
\noindent
where $\alpha$ is an analyzing power related to the spin of the
final state ($\alpha$=1 for $\pi\nu$ mode and $\alpha$=0.46 for $\rho\nu$ mode).
Figure~3c) shows the observed spectra for $\pi\nu$ mode.
The smaller analyzing power for the $\rho\nu$ mode, which comes from the mixture of opposite polarization states of $\rho$, can be restored by introducing a
second kinematical variable \cite{spin} to discriminate the polarization state.
In the present analysis, we used the decay angle of $\rho$ in the $\tau$ rest
frame relative to the $\tau$ flight direction ($\cos\psi_{\tau}\approx 2x-1$) and the decay angle of the charged $\pi$ in the $\rho$ rest frame relative to the $\rho$ flight direction
(cos$\psi_{\rho}$).
The  $\cos\psi_{\rho}$ distribution was examined in two $\cos\psi_{\tau}$ regions, $\cos\psi_{\tau} < 0$ and $\cos\psi_{\tau} > 0$, separately.
 The observed distributions are shown in Fig.~4.

The polarization was evaluated by fitting the linear combination of the kinematical
distributions expected for the full polarization cases ($h=\pm1$)
to the corresponding observed distributions.
The fit was done by minimizing a $\chi^2$ defined as
\begin{align}
\chi^2 &= \sum_i \left[ \frac{N_i^{\rm meas}-\frac{1}{2}[(1+\ptau)(N_i^+ + N_i^{\rm BG +})+(1-\ptau)(N_i^- + N_i^{\rm BG -})]}{\sigma_i} \right]^2,
\label{kai}
\end{align}
where $N_i^{\rm meas}$ is the number of entries in $i$-th bin, $N_i^+(N_i^-)$ and $N_i^{\rm BG +}(N_i^{\rm BG -})$ are those of 
Monte Carlo events with $h=+1$($-1$) for the signal and the background coming 
from other decay modes of $\tau$, respectively. The denominator, $\sigma_i$ is 
the statistical error in each bin. The expected contributions from
non-$\tau$ process were subtracted from $N_i^{\rm meas}$ beforehand. 

The Monte Carlo events for each helicity state of $\tau$ leptons were generated 
by using
KORALZ 4.0 \cite{KORZ} for $\tau$ pair production and TAUOLA2.5 \cite{TAUOLA}
for $\tau$ decay. The radiative effect in the initial state, which affects the energy spectra substantially, is properly incorporated up to $O(\alpha^2)$ in the event generation.
The efficiencies in the decay mode identifications were treated according to the functions evaluated in the previous section.
 The results of the fit are shown in Figs.~3 and 4 with histograms.
The average polarizations of $\tau$ thus evaluated are
tabulated in Table~ 2. 

The forward-backward asymmetry of the polarization, \ap, was evaluated by the same procedure as
above in the forward ($0 \leq \cos\theta < 0.5$) and backward ($-0.5 < \cos\theta < 0$)
region, separately. The scattering angle
was determined by the thrust axis of the event and the charge of the
daughter particle.
 From the forward and backward polarization thus evaluated, its
asymmetry was calculated as listed in Table 3.

The systematic errors in the above measurements were considered in three
categories.

\noindent {\underline {Detection efficiency}}: 
The evaluation of polarization can be affected by uncertainties in the momentum dependence of the efficiency of the decay mode identification. 
Since the estimation was made
using real data, there are substantial statistical 
uncertainties. Their effects on $\ptau$ were estimated by changing the
efficiency independently in each momentum bin within their errors. 

\noindent {\underline  {Background}}: Uncertainties related to the
background, either from mis-identified $\tau$ decays or non-$\tau$ processes,
can lead to the systematic error.
The evaluation were found to be very stable with the possible variation of the
background fraction.

\noindent {\underline {Monte Carlo statistics}}: Because of limited
statistics in the Monte Carlo simulation, the generated spectra had some
uncertainties. The systematic errors due to them were estimated by changing
the number of entries in each bin of the spectra within their statistical errors.

\noindent The systematic errors thus estimated are summarized in Table 2 and Table 3.
As found in the tables, the systematic errors are significantly smaller
than the statistical ones in the present analysis. 

Combining the results for
all the modes studied in the present analysis with neglecting the
small systematic errors safely,
the average polarization of $\tau$ leptons and its forward-backward asymmetry
are 
\begin{align}
\ptau= 0.012\pm0.058, \\
\ap = 0.029\pm0.057.
\end{align} 
\noindent Although the present
results are consistent with zero and are also consistent with the values
predicted by the Standard Model (0.028 and 0.021, respectively, 
at $\sqrt{s}=$ 58 GeV),
they are the first substantial measurements in the energy region below the $Z^0$ pole as found in Fig.~5, where the results from the present and the previous 
experiments are plotted together with the standard model prediction.

\section{Conclusions}
The polarization of $\tau$ leptons and its forward-backward asymmetry in the reaction $\process$  have 
been measured at the center-of-mass energy of 58 GeV. They are evaluated to be 
  $0.012\pm0.058$ and $0.029\pm0.057$, respectively, combining the results for all the decay modes studied. The prediction of the standard model is confirmed in the energy region below the $Z^0$ pole.

\vspace{10mm}
\newpage
\noindent {\Large \bf Acknowledgment}

We wish to thank the TRISTAN machine group for their patient efforts regarding the accelerator
operation that continued for many years. We gratefully acknowledge the outstanding contributions
of the technical staff at KEK and the collaborating institutes who participated in the construction
and operation of the VENUS detector. The data acquisition and analyses were made possible with
continuous support by people from the on-line group and the computer center of KEK.
We thank K.~ Hagiwara for useful discussions. Thanks also go to Z.~ Was for his help to use
the KORALZ and the TAUOLA Monte Carlo generators.

\newpage

\begin{table}[H]
\caption{The number of identified decays, the estimated background contribution 
and the selection efficiency for each decay mode. Also listed are calculated branching ratios and corresponding PDG96 values.}
\label{BR}
\vspace{5mm}
\begin{tabular}{lcccc}
\hline
      & $\rho\nu$ & $\pi(K)\nu$ & $e\nu\bar{\nu}$ & $\mu\nu\bar{\nu}$ \\
\hline
\# observed   & $829$ & $287$ & $564$ & $628$ \\
\hline
\# background($\tau$) & $75.6\pm8.1$ & $25.9\pm1.6$ & $2.0\pm0.4$ & $6.2\pm0.8$ \\
\# background(non-$\tau$) & $17.7\pm2.4$ & $8.7\pm1.6$ & $20.3\pm4.0$ & $21.7\pm1.9$ \\
\hline
Efficiency(\%)&$36.1\pm1.0$ & $24.3\pm0.8$ & $34.5\pm1.0$ & $41.3\pm1.3$ \\
\hline
BR(\%)       & $24.1\pm0.9\pm0.8$ & $12.3\pm0.8\pm0.5$ & $18.6\pm0.8\pm0.7$ & $17.2\pm0.7\pm0.6$ \\
PDG96(\%)    & $25.24\pm0.16$ & $12.03\pm0.14$ & $17.83\pm0.08$ & $17.35\pm0.10$ \\
\hline
\end{tabular}
\end{table}

\vspace{10mm}

\begin{center}
 \begin{table}[H]
 \caption{Evaluated $\ptau$ and systematic errors contributing to it. Errors indicated in the
first column represent statistical ones.}
\vspace{5mm} 
  \begin{tabular}{lcccc}
   \hline
                 &$e\bar{\nu}\nu$&$\mu\bar{\nu}\nu$&$\pi(K)\nu$&$\rho\nu$  \\
   \hline
   $\ptau$	 &$0.13\pm0.18$&$0.15\pm0.20$&$-0.01\pm0.12$&$-0.03\pm0.07$ \\
   \hline\hline
   Efficiency    &0.021&0.021&0.018&0.024 \\
   $\tau$ BG     &0.014&0.013&0.010&0.009 \\
   Non-$\tau$ BG &0.012&0.008&0.009&0.011 \\
   Monte Carlo statistics.&0.021&0.020&0.013&0.015 \\
   \hline
   Total systematic error&0.035&0.033&0.026&0.032 \\ \hline
  \end{tabular}
 \label{sys_err on ptau}
 \end{table}
\end{center}

\begin{center}
 \begin{table}[H]
  \caption{Evaluated $\ap$ and systematic errors contributing to it. Errors indicated in the
first column represent statistical ones.}
\vspace{5mm}  
\begin{tabular}{lcccc}
   \hline
	         &$e\bar{\nu}\nu$&$\mu\bar{\nu}\nu$&$\pi(K)\nu$&$\rho\nu$  \\
   \hline
   $\ap$	 &$-0.02\pm0.17$&$-0.01\pm0.19$&$0.03\pm0.12$&$0.04\pm0.07$ \\
   \hline\hline
   Efficiency    &0.023&0.025&0.017&0.026 \\
   $\tau$ BG     &0.015&0.012&0.011&0.009 \\
   Non-$\tau$ BG &0.013&0.008&0.010&0.014 \\
   Monte Carlo statistics.&0.022&0.019&0.015&0.013 \\
   \hline
   Total systematic error&0.038&0.035&0.027&0.033 \\ \hline
 \label{sys_err on ap}
  \end{tabular}
 \end{table}
\end{center}

\newpage

\begin{figure}[H]
\includegraphics[scale=1.0]{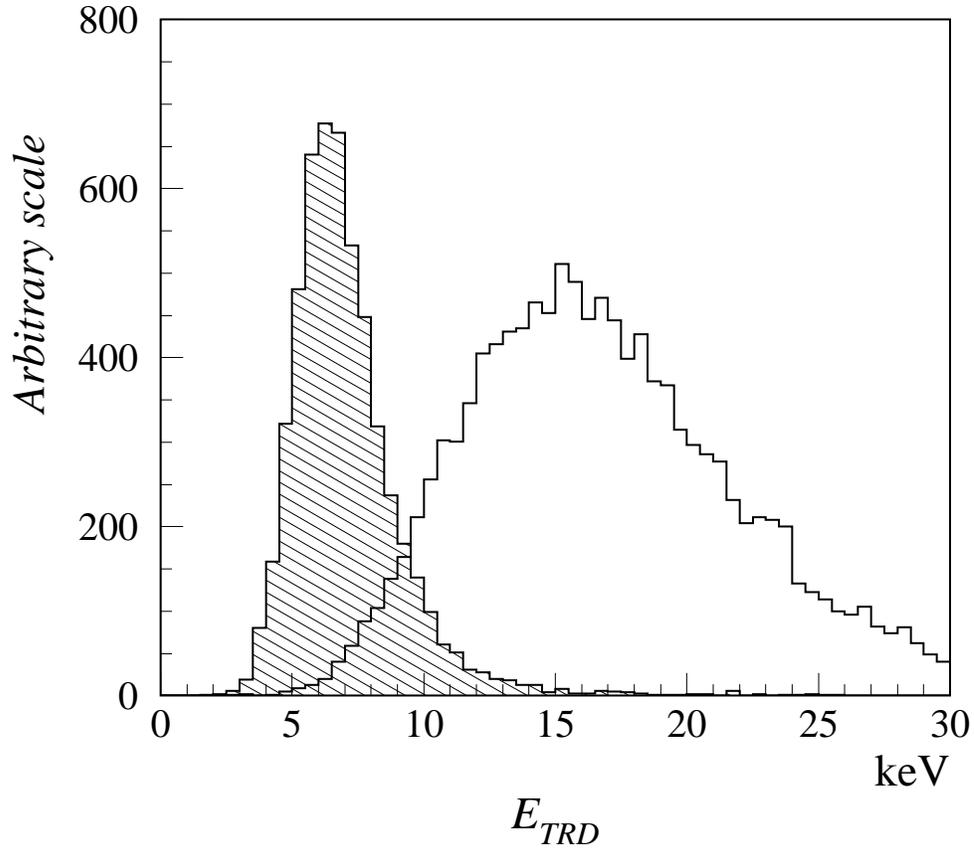}
\caption{Energy deposit spectra in the TRD, for electrons (solid line histogram) and for muons 
(hatched histogram) with their momenta greater than 2 GeV/$c$.
The spectrum for pions must be the same as that for muons. 
The cut at 10 keV provides a clear discrimination between electrons and pions.}
\label{trd spectra}
\end{figure}

\newpage

\begin{figure}[H]
\includegraphics[scale=0.7]{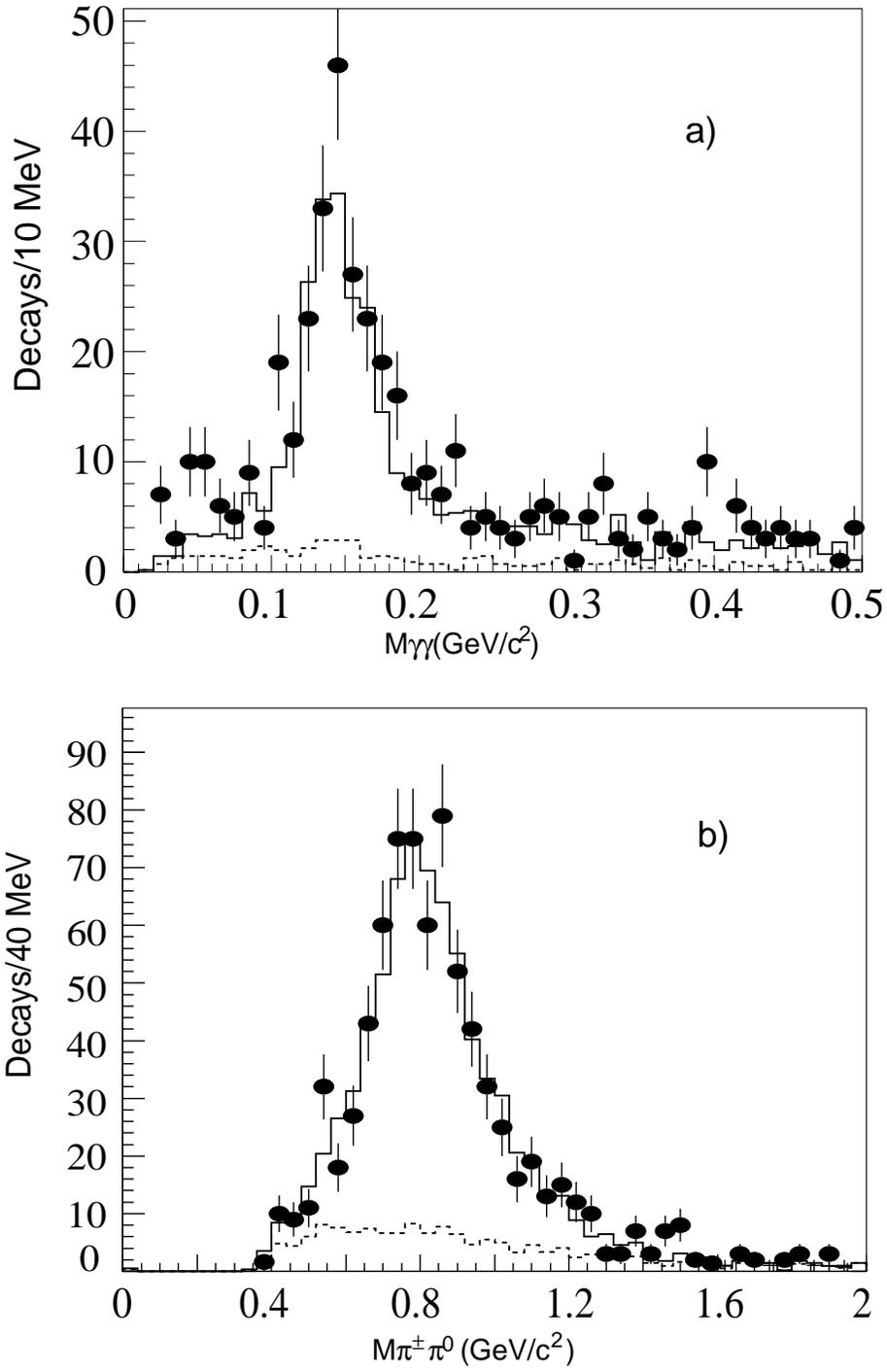}
\caption{Invariant mass distribution of a)the two photons and b)$\pi^{\pm}\pi^0$ in the selection of the $\rho\nu$ decay mode.
The solid histogram shows the
simulation result. The estimated background contributions are indicated with
the dashed histogram.
}
\label{mass distribution pi}
\end{figure}

\newpage

 \begin{figure}[H]
 \includegraphics[scale=0.8]{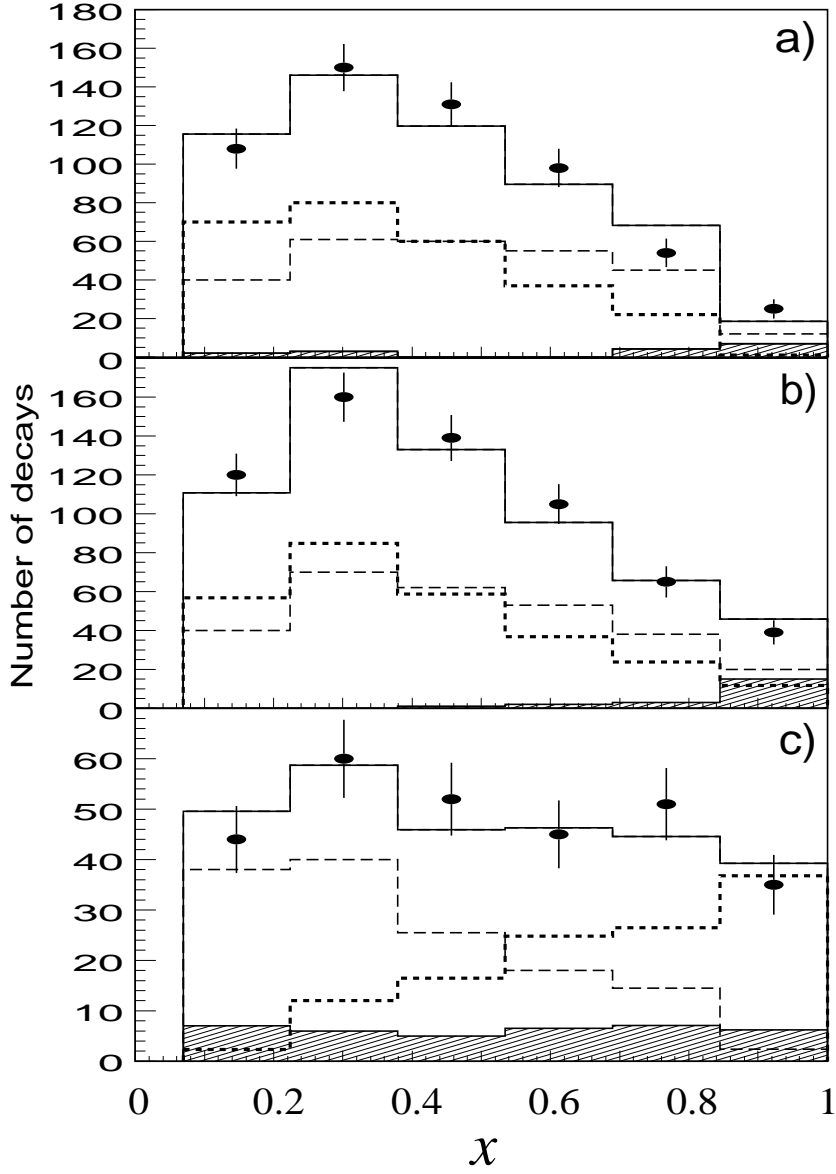}
\caption{Scaled energy distribution for a)$\tau\rightarrow e\nu\bar{\nu}$, b)$\mu\nu\bar{\nu}$ and c)$\pi\nu$ modes.
The filled circles indicate the selected candidates. 
The solid line histograms represent
the result of the fit. Contributions from each helicity state are
indicated by the dashed ($h=+1$) and dotted ($h=-1$) histograms.
The background contributions are illustrated with the shaded histogram.
} 
\label{x fit}
\end{figure}

\newpage

\begin{figure}[H]
  \includegraphics[scale=0.8]{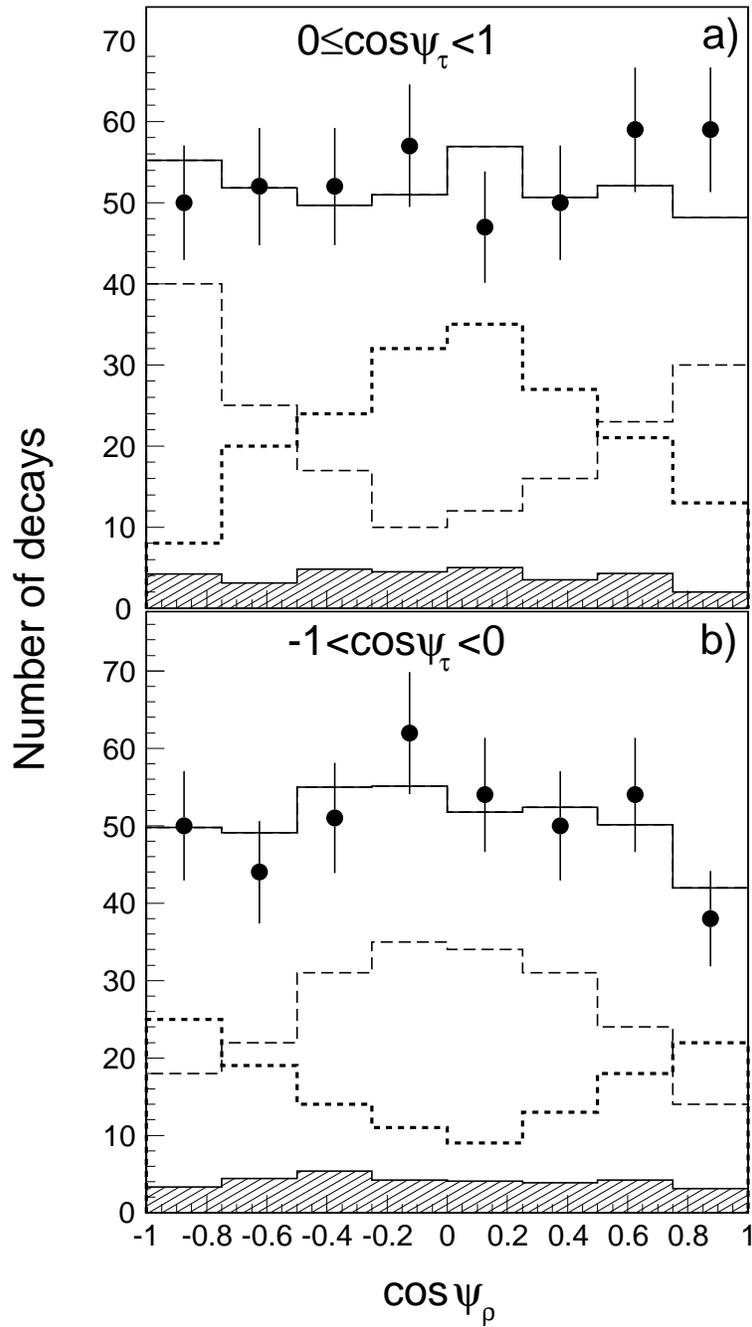}
\caption{Observed $\cos\psi_{\rho}$ distributions for
the $\tau\rightarrow\rho\nu$ candidates in the region of
a)$0<\cos\psi_{\tau}$ and b)$\cos\psi_{\tau}<0$.
The filled circles indicate the selected candidate.
The solid line histograms represent the result of the fit.
Contributions from each helicity state are indicated by the dashed ($h=+1$)
and dotted ($h=-1$) histograms.
The background contributions are illustrated with the shaded histogram.
} 
\label{rho fit1}
\end{figure}

\newpage

\begin{center}
\begin{figure}[H]
 \includegraphics[scale=0.9]{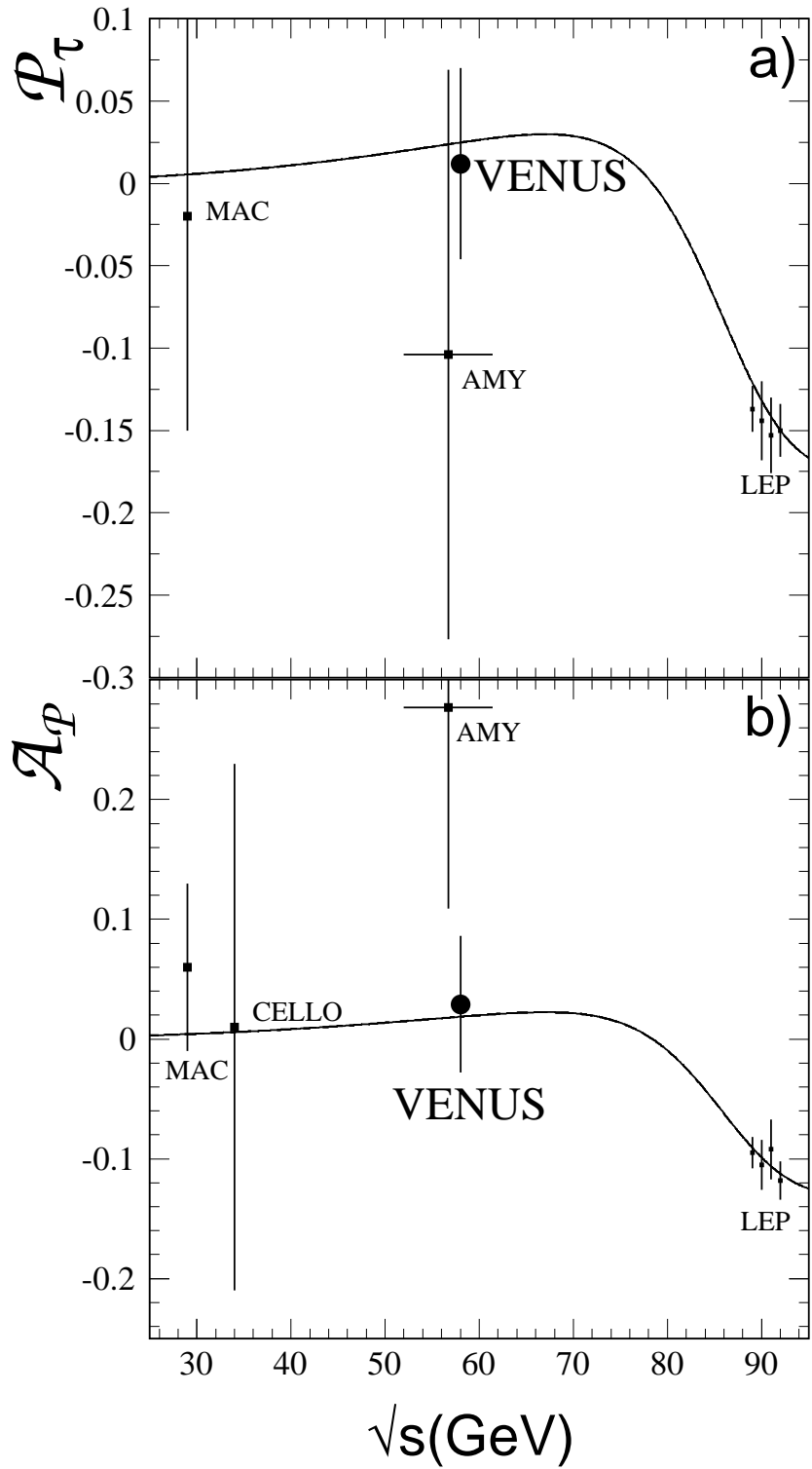}
\caption{Measured a)average polarization $\ptau$ and b)polarization asymmetry 
$\ap$ together with those from other
experiments. Solid line indicates the
Standard Model prediction.
}
\label{results with other experiments}
\end{figure}
\end{center}

\end{document}